# Experimental Indications of Electro-Gravity

T. Datta, Ming Yin[1], Andreea Dimofte, M. C. Bleiweiss[2] and Zhihua Cai

Physics & Astronomy

University of South Carolina, Columbia, SC 29208

1) Benedict College, Columbia, SC 29204

2) Navy Academy Preparatory School, Newport, RI 02841

Correspondence and request for materials should be addressed to Timir Datta (email:datta@sc.edu)

**Abstract:**

Recent results from our on going experimental investigation of the influence of space dependant electric fields on the weight of test particles are reported. Test particles were gold coated metal spheres of same size but of different masses. Data collected from a number of runs over several years continue to indicate an intriguing effect. For experimental parameters in question this effect is manifested as a ppm level sample mass dependent force additional to expected electrostatic forces. A force that is proportional to mass is the unique signature of gravity furthermore it is non-zero only when the field is applied; hence these observations may be further evidence in support of electro-gravity.

**Background:**

Discovering commonality amongst disparate phenomena is a raison d'etre of science. For example in 1831, Michael Faraday[1] provided the foundation for electromagnetism. These results[2] led James Clerk Maxwell[3] (1873) to further merge electrodynamics with optics. Faraday was also amongst the first to explore electro-gravity effects but obtained null results. Led by Einstein and Cartan[4], theoretical unification of gravitation with electromagnetism is being pursued from the 1920's. Since the late 1990's in our laboratory we have been seeking experimental evidence for electro-gravity. Here we provide some recent tantalizing data on a mass dependant force observed in inhomogeneous electric fields. If confirmed this will evidence of unification of electromagnetism with gravitation.

**Experimental:**

The goal of this experiment[5-7] has been to investigate electrical influence on gravity. The presence of the electric field (**E**) considerably increases the complexity of our apparatus as incorporates both gravity and electromagnetic aspects. The apparatus was kept minimal so that the results can be as transparent as possible. The gravitational role was assigned to a collection of test particles (masses). These particles were metal spheres of same shape and size but of different chemical compositions (industrial grade aluminum, steel etc) and different densities, hence different masses (m). A large range in the masses is needed so that mass dependence of the signal could be tested hence we added several tungsten carbide spheres to our collection. The test masses were stored in static free conditions and grounded to the earth so that electrical charge neutrality can be maintained as far as possible. Also to ensure uniformity in electrical and surface properties, the test masses and electrodes in the apparatus were sputter coated with a thin film of gold.

The principal components of our apparatus (Fig. 1) were (i) an electronic balance (ii) a measurement cell containing the "Stern-Gerlach" capacitor to be descried later and, (iii) a computer. At first the whole apparatus was setup in an underground chamber below a large climate controlled laboratory where mechanical vibration and electrical noises were low. Later on the apparatus was moved to the fifth floor of the physical science center building[8].

The test mass was suspended with an electrically insulating string from the balance and placed at a desired position between the top flat and the conical bottom electrodes of the capacitor (marked 2 in figure 1) and the measurement cell was filled with high purity nitrogen gas. The net charge collected due to electric discharges during measurements was below $10^{-11}$ coulombs[7]. To keep the balance condition same each test particle was measured at constant total weight by placing (away from the electric field) additional compensating ballast weights (marked 3 in figure 1).

We reasoned that to produce electro-gravitational influences uniform, static fields have to be avoided. Generally, if changes in both space and time variables are relevant then dimensional analysis imply that in the first order the space dependant term would be a factor of c times that of time dependent term, where c is the speed of light. Hence we chose to employ just a spatially non-uniform field under nearly stationary conditions and designed a capacitor with a conical electrode or an electrical analog of the Stern-Gerlach system to create such a field. The electrostatic field produced by this capacitor was cylindrically symmetric and along the central axis, a function only of the vertical position, z.

Every mass was positioned at the same location and was subjected to identical voltage time sequence or pulses. Each measurement comprised of several fifty second long segments of zero, positive, zero, negative and zero voltage pulses applied between the electrodes. The voltages were ramped up and down to the desired values in a few seconds. The periods of zero voltage maintained before and after each pulse were for baseline purposes.

The weight force on the masses was measured by the digital balance as a function of time while the bias voltage V(t) cycle described above was applied between the capacitor electrodes. The weights ($W_+$, $W_0$ and $W_-$) or balance output under the applied voltages (+V, 0 and -V) were recorded by the computer. The force measurement had a resolution of ~100 ppb for the largest test masses.

Figure 2 shows the balance output (W) and the applied voltage (V) as functions of time. One cycle of the staggered square-wave-like positive and negative (2kV) pulses and the intervening periods of zero bias are plotted in the upper graph. The lower curve shows the corresponding balance out put. As can be seen when the applied bias is zero the balance output is $W_0$ which is the "free weight" of the test particle. When a bias is applied to the capacitor electrodes the measured weight is higher than $W_0$ indicating an increase in the total downward force on the test mass so the balance output is increased. Let $W_+$ and $W_-$ be the balance outputs with positive and negative biases applied respectively. Under positive bias the weight increase ($W_+ - W_0$) is larger than that for negative bias. Or in other words by manipulating the bias one may control or influence the earth's gravitational pull on the test mass.

In this publication we present results with 1kV pulses, from a batch of ten test masses (19.115± 0.096 mm diameter spheres) measured in the fifth floor lab, on eleven different days (runs) during the 2003-2004 winter break.

**Sources of Errors:**

In light of past controversial observations in gravity [9] one can never be too careful when dealing with intriguing effects in gravitation. Hence we have tried to correct for the obvious experimental artifacts also a compromise was made between obtaining enough statistics and "signal drift" associated with long time intervals.

Some of the difficulties that we encountered are typical of gravitational experiments[10-14]; namely, the signal of interest is small, in the ppm range, background noise, large drift and the problem of shielding stray influences. Noise due to mechanical vibrations and air currents were reduced by seclusion, by vibration isolation and by enclosing the test rig inside of a thermally insulated and environment controlled glove box. To minimize systematic errors, the samples were selected randomly. Each mass was tested over two or three runs in succession. Also we monitored the stretch in the suspension fiber as well as any significant deviations from the chosen position of the sample. No difference was observed amongst the different masses.

Furthermore, the apparatus was simultaneously monitored by a video camera and a number of sensors[7]. For example, time stamped data for temperature, barometric pressure, ambient electric field and the acceleration of the base of the balance were recorded and analyzed. Strong correlations were observed between air temperature and the balance output. Typically, as temperature increased there was a decrease in the balance output at the rate of ~1 μg/K. Generally the measurements were performed with temperature excursions~1K/day. Weaker perturbations were noted for the other parameters. Such as, white featureless noise outputs of the accelerometer indicated that the apparatus was free of sudden impacts.

In addition to the random noise, drifts and cyclic behaviors were noticed in the weight signal. We have identified the major component of the time dependant changes to be associated with the solar and lunar tides[15]. Tides have both long time and shorter time correlations. Monthly periods in the tides affect gravitational experiments and are seen in published data[16]. Diurnal tidal variations of shorter periods may also be seen in the gravitational background. Fig. 3 shows the correlation between the tide and the balance out put for a 50 g test mass. Between high and low tides over short periods of time, the apparent weight of the test mass can change at the rate of ~1 ppb/min. Figure 3 also demonstrates the quality and fidelity of our measurements. Generally, uncontrollable change in the background is not a "good thing" but the tides being so regular provide excellent calibration points and give additional validation to the measurements of the weight force. To avoid systematic diurnal influences, the experiments were performed at different times and different mass order. Also records were kept of the laboratory humidity and temperature.

**Results:**

To better understand our observations we analyzed the difference

$$\Delta = W_+ - W_- . \qquad \ldots (1)$$

Where, as defined earlier $W_+$ and $W_-$ are the balance outputs with positive and negative biases applied respectively. By definition $\Delta$ is zero in zero field (V=0). The noise was at the level of instrumental resolution ~10 nN where as the $\Delta$'s are ~$\mu$N, i.e., the observed signal is two orders of magnitude larger than the noise. However, the error bars for the experiment as a whole depend on charging, drift and run-to-run scatter.

Eleven experimental runs, all with 1kV pulses were taken on different days between 17, Dec'03 and 23, Jan'04. For each run, different numbers of samples out of ten (19 mm) test particles were measured in random order of their masses. The graphs of the $\Delta$, versus the test particle mass, $W_0/g$, (g is the local free-fall acceleration due to gravity) obtained from the data of five sets of runs out of the eleven sets are shown Figure 4. The straight lines are the least square fits for each day. Note that from day to day and even on a single day the points show considerable scatter; however, even with this noise a general the signal tends to increase with increasing sample mass.

The linear fits of the graphs reveal a contribution which directly correlates with the mass of the sample. $\Delta$ in equation 1 can be further decomposed as follows:

$$\Delta(m) = \Delta_{EM} + \Delta(m)_{EG} \qquad \ldots (2)$$

In equation 2, $\Delta_{EM}$ is the y-intercept of the graph. It is also the force in the limit of zero sample mass and can be attributed to pure electromagnetic effects. Equation 2 may also be written as,

$$\Delta(m) = \Delta_0 + \alpha * m \qquad \ldots (3)$$

The positive slope observed of the experimental data in the figure indicates that the $\Delta_{EG}$ term in equation 2 is directly proportional to mass. That is the slope (d$\Delta$/dm) or $\alpha$ in equation 3 is positive. The observed scatter in the data gives rise to changes in both the slope and the intercept. Scatter appear from a number of sources of noise introduced in the signal namely mechanical vibration, corona charging and variations in dielectric properties in the cell due to variations in the fill-gas composition.

For our experimental parameters from the five sets of runs with these ten test masses we find $\Delta/W_0$ is in the ppm range and $\alpha$ is of the order of $\mu N/kg$. The run dates, values of $\Delta_0$, $\alpha$ from equation 3 and the fit correlation coefficient ( R ) are given in table 1. The negative value of $\Delta_0$ implies that the net average charge acquired by the samples in a particular run was negative. Charging takes place mostly during the discharge, this is one reason why we reduced the pulsing voltage to 1kV for this set of runs. The best fit of all the data from the eleven sets taken as one, gives $\Delta_0$ equal to 0.78 $\mu N$ and $\alpha$ equal to 48.7 ($\mu N/kg$). In comparison in 2001-2002, $\alpha$ was found to be 63.9 ($\mu N/kg$) from data taken with 2kV pulses, at the better underground lab, but without tungsten carbide samples and not all under constant balance loading. It is unlikely that it is just a random coincidence that the slopes from different sets of runs in two locations with new samples will be so consistent.

Large gradients of E fields are not common in nature except in small systems. In condensed systems such as crystalline solids particle dynamics is dominated by the diffractive properties of the quantum wave function which gives rise to large, small, or even negative effective mass. Experiments on gravitational behavior in atomic or subatomic states may be interesting. But our search[17] on test particle behaviors in electric fields indicates that the presence of such a novel force may be consistent with the data available in the printed literature.

In contemporary paradigm, non-general relativistic, unification of electromagnetism with gravity and at the low energy scales of our experiment might amount to heterodoxy. We are not aware of any accepted mechanisms with falsifiable, quantitative predictions for our experiment to be tested against. However, a novel theory of gravitation proposed by Vargas and Torr[19] is consistent with inhomogeneous electric fields generating gravitational forces.

**Summery:**

The test particles are of different mass but of the same shape, size and similar electrical properties. The signal ($\Delta$) is non-zero only when the inhomogeneous electromagnetic field is applied. $\Delta$ is sample mass dependent. Also we observed that the smaller the sample size the larger is the effect, which may be another indication of the role of field inhomogeneity. These observations imply an intimate connection between the test particle-earth gravitational field and electric field.

The accumulated data spread over several years and performed at two locations indicate the presence of a novel force that persists above the experimental scatter. For our setup, it is a ppm level force ($\Delta_{EG}$) which scales with sample mass but disappears when the spatially inhomogeneous electric field is turned off; hence likely to be an electro-gravity signal.

Such electro-gravity is non-general relativistic. However, as we have recently reported[20] electro-gravity is not inconsistent with the experiments by Ehrenhaft[21] and LaRue[22] on charges and electric fields. Furthermore, it is might not be a coincidence that electromagnetism is manifested in presence of time dependence and the present observations concerns gravitation in spatially inhomogeneous electric fields. As mentioned at the beginning, Faraday was one of the pioneers to seek connections between electricity and gravity; but, did not use inhomogenous fields and obtained no effect. He ended his landmark 1851 paper (Phil. Trans Royal Soc. pp1-122) with the following classic words[18] "…Here end my trials for the present. The results are negative; they do not shake my strong feeling of an existence of a relation between gravity and electricity…".

**Acknowledgements:**

The authors express their thanks to Anca Lungu, James Jones and R. Leonard for assistance at various stages of the measurements. Sir John Meurig Thomas for reminding us of Faraday's pioneering contributions[18]. Jose Vargas for bringing his theory[19] to our attention and for continued interest in this project.

**Table 1**

| Date | $\Delta_0(\mu N)$ | $\alpha\ (\mu N/kg)$ | Correlation ( R ) |
|---|---|---|---|
| 18Dec'03 | -1.16 | 36.2 | 0.32 |
| 26Dec'03 | 0.94 | 85.9 | 0.93 |
| 2Jan'04 | 16.7 | 29.2 | 0.43 |
| 12Jan'04 | .50 | 58.5 | 0.58 |
| 23Jan'04 | -.90 | 172.0 | 0.94 |
| Mean all runs | .78 | 48.7 | 0.80 |

**Figures with captions:**

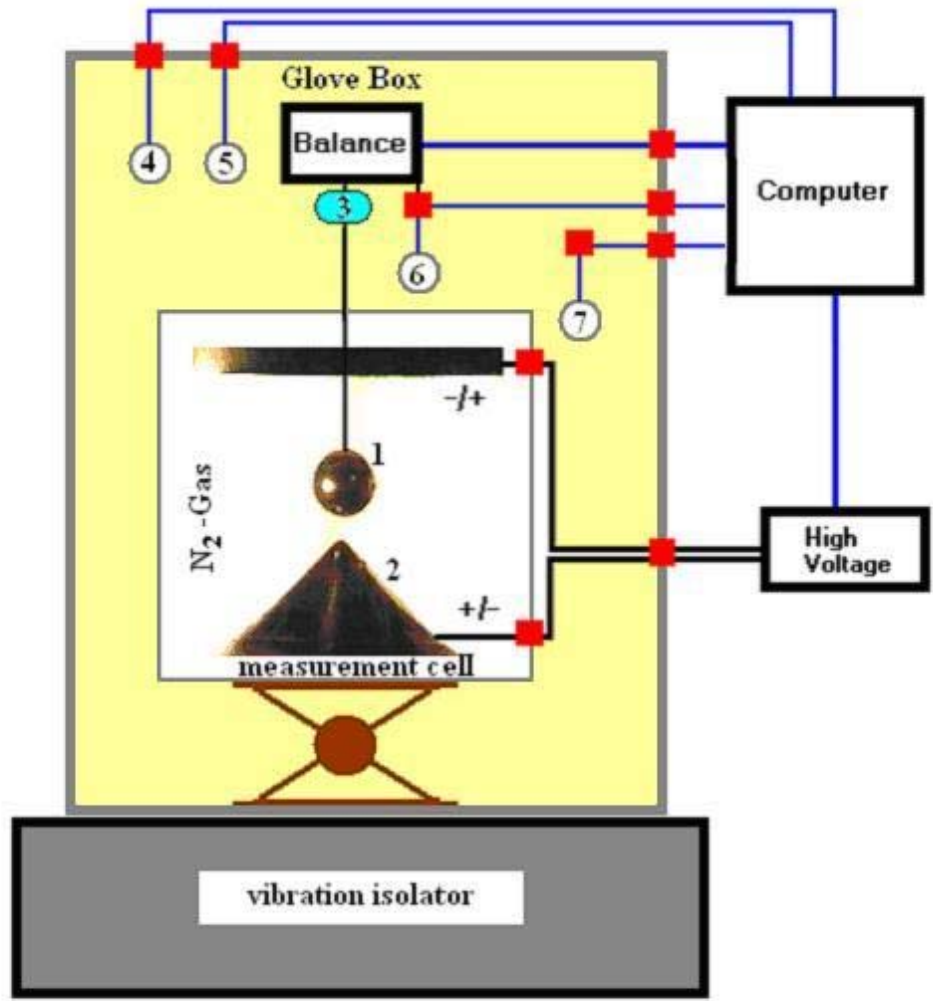

**FIGURE 1:** A sketch of the apparatus showing a test mass (1) hanging in the inhomogeneous field between the flat and the conical (2) electrodes. The ballast weight (3) and temperature sensor (4), barometric pressure sensor (5), accelerometer (6) the electrometer probe (7) and some major equipment are also indicated.

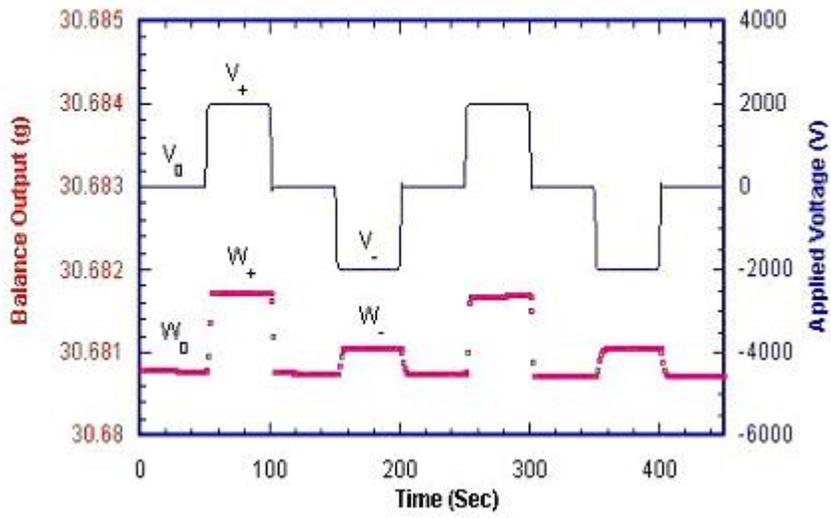

**FIGURE 2:** Correlation between a 2kV applied bias pulse on the electrodes and the balance output as functions of time.

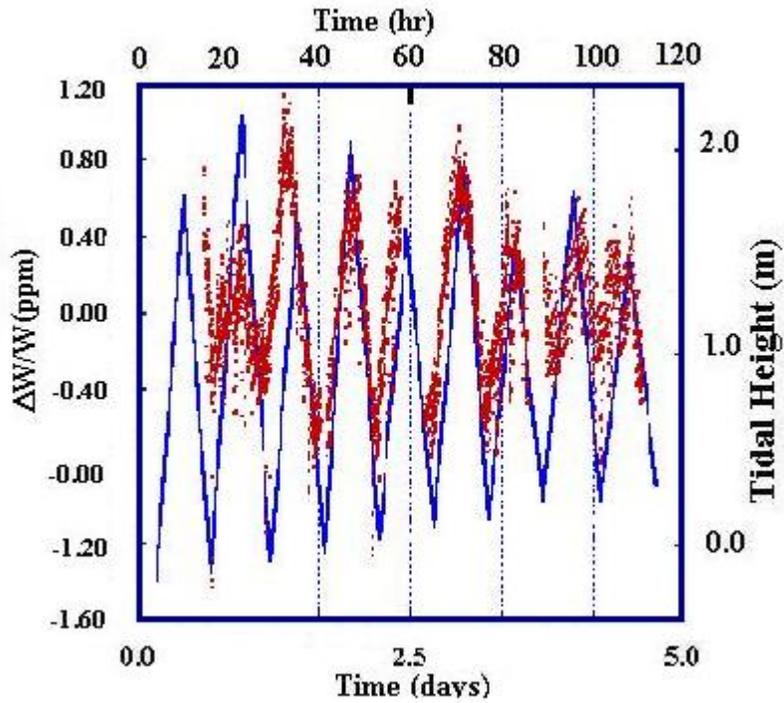

**FIGURE 3:** Scatter plot of our weight data for a fifty gram standard vs. time, compared with NOAA table values of the high and low tides at Charleston, SC. The correlation between weight and daily tides indicated by the interpolated triangular solid graph is excellent. This chart reflects the high fidelity of the balance output as well as the overall quality of our data analyses.

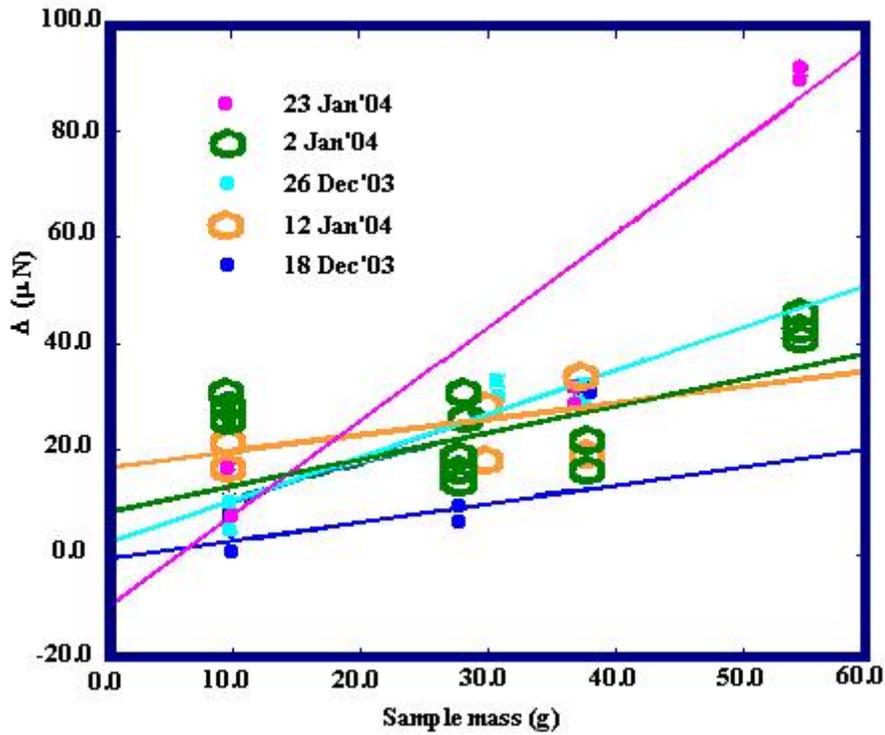

**FIGURE 4:** Δ versus sample mass, results from the "fifth floor" lab for ten 19mm test masses. Data taken on five different days at 1kv bias pulses are shown. To reduce overlap not all the data are marked and some others are slightly staggered on these graphs. The straight lines are the least squares best fits for each day.